\documentclass[ 
preprint, 
superscriptaddress,
 amsmath,amssymb,
 aps, physrev,
]{revtex4-2}

\usepackage{graphicx}
\usepackage{dcolumn}
\usepackage{bm}
\usepackage[normalem]{ulem}
\usepackage{xcolor} 

\newcommand{\utah}{\affiliation{Department of Chemistry,
  The University of Utah, Salt Lake City, Utah, United States}}

\begin{document}


\title{\textbf{Bridging simulation length scales with cellular automata} 
}%

\author{John J. Karnes}
 \email{Contact author: karnes@llnl.gov}
 \affiliation{Lawrence Livermore National Laboratory,
   Livermore, California, United States}
\author{Esteban D. Gadea}
 \utah
\author{Shakkira Erimban}
 \utah
\author{Ignacio J. Bombau}
 \utah
 \affiliation{INQUIMAE, Universidad de Buenos Aires, Buenos Aires, Argentina}
\author{Valeria Molinero}
 \utah

\date{\today}

\begin{abstract}

Multiscale simulation requires coupling physics models that operate at different characteristic length and time scales, because no single method spans the range needed for most problems of interest. 
Moving to a coarser-grained representation unlocks longer length and time scales, but it discards the microscopic interactions that build morphology. A fine-grained model can be initialized from an arbitrary packing and left to self-assemble into a physically meaningful structure; a lower-resolution model cannot, and must inherit its starting configuration from a higher-fidelity simulation. The length scales accessible to the coarse-grained model are therefore set not by the coarse-grained method itself, but by the largest fine-grained configuration that can be affordably equilibrated.
A representative example is the scale-up from particle-based molecular dynamics (MD) to a lattice-based representation such as kinetic Monte Carlo (kMC). In this work, we present a cellular automata (CA) approach for generating arbitrarily large lattice starting configurations. CA is a natural fit for this task: short-ranged local rules drive the evolution of a lattice, and their repeated application gives rise to emergent long-range order, thematically mirroring how short-ranged interactions in MD produce self-assembled morphology. We use a configuration from a higher-fidelity simulation as a training set and learn the CA rules from it via logistic regression. As a proof of principle, we develop these rules for a hydrated anion exchange membrane (AEM), generate new starting configurations, and benchmark their performance in mesoscale kMC simulations against an MD-derived ``ground truth.'' We then demonstrate the ability to generate substantially larger lattices and show that their behavior in kMC is consistent with that of the smaller CA benchmark configurations.

\end{abstract}

\maketitle



\section{Introduction}

Pairing theory with modern computational resources, computer simulation has established itself as a core component of contemporary research in the physical sciences. The nature of the system being studied and insights sought by researchers guide the design of simulation studies, where model fidelity and computational expense must be balanced. A grand challenge in simulation work is the bridging of length and time scales across different simulations of the same system~\cite{karplus_MolecularDynamicsSimulations_2002,voth_CoarseGrainingCondensedPhase_2008}, particularly when researchers use the results of high-fidelity simulations to develop models for larger simulations~\cite{noid_MultiscaleCoarsegrainingMethod_2008}. Reduction in compute expense is required to access these larger scales, which generally implies the use of a model with an entirely different formalism for both the spatial representation of the components and describing their evolution through time. For example, while \textit{ab initio} simulations are limited to a few hundred atoms and sub-nanosecond timescales, classical molecular dynamics simulations represent atoms by a simpler model and routinely capture hundred nanosecond trajectories with millions of atoms in a simulation cell~\cite{allen_ComputerSimulationLiquids_1987}. In addition to developing a new, appropriate physics model, simulations of larger scale require a larger starting configuration. This aspect of scale-up is not trivial when considering systems where the behavior of interest is a function of microscopic morphology. In this work we introduce a general approach and philosophy toward the generation of these larger starting configurations, required when attempting to `scale up' simulations and implement less computationally expensive methods. 

\begin{figure}[h]
\includegraphics[width=3.25in]{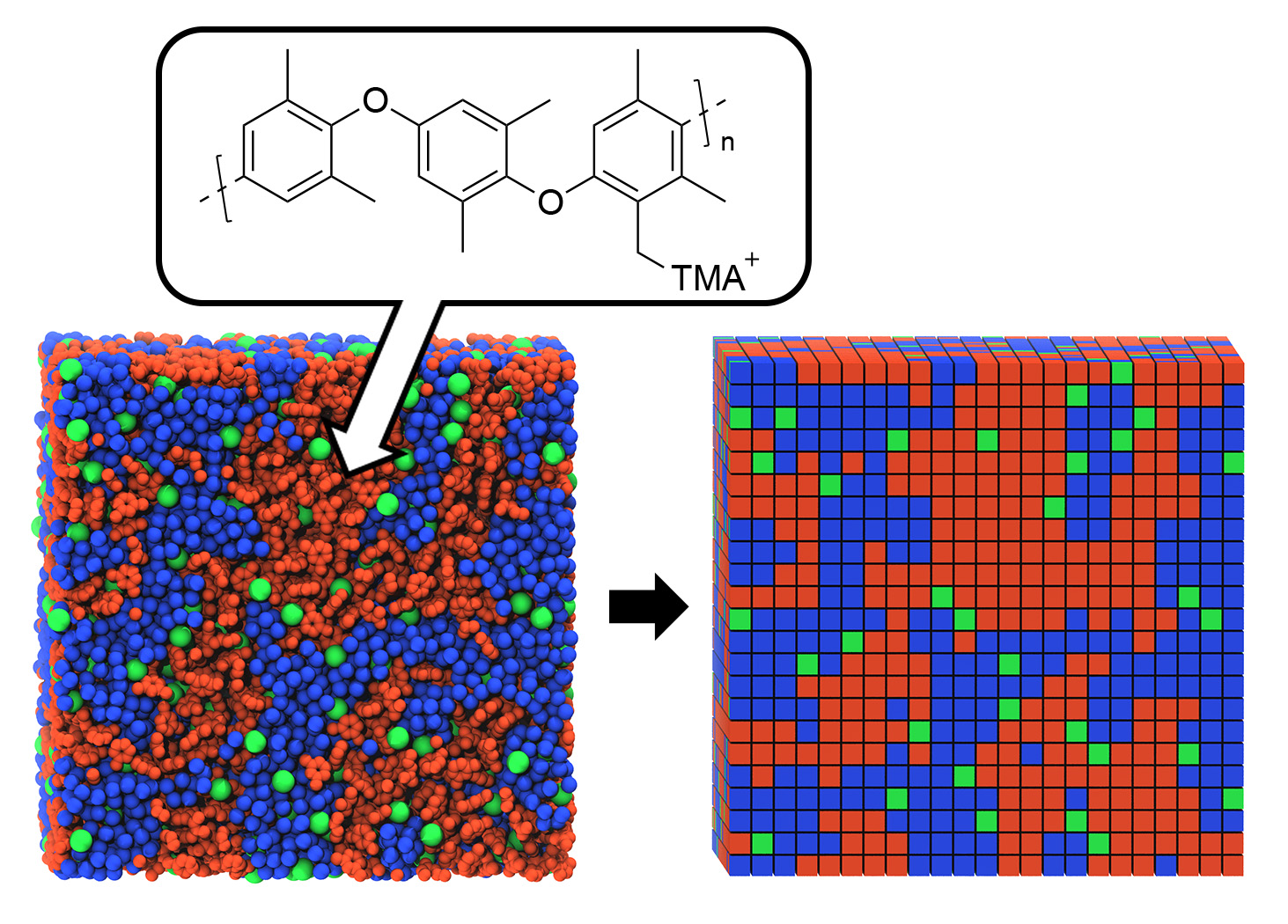}
\caption{\label{fig:boxes} Inset, top: Structure of the PPO-TMA polyelectrolyte. MD configuration (left) and lattice representation (right) of hydrated PPO-TMA, where polymer is red, water is blue, and cation sites are green.}
\end{figure}

As an exemplar computational framework, we consider the scale up from classical molecular dynamics (MD) to a lattice-based kinetic Monte Carlo (kMC) simulation, shown in Figure ~\ref{fig:boxes}~\cite{gadea_KineticMonteCarlo_2026}. This is a representative case where a high-fidelity off-lattice, particle-based simulation is used to define the starting configuration for a mesh- or grid-based representation. In this mesoscale regime, the atomistic description of the physical system used in MD is abandoned for a lattice of cells~\cite{voter_INTRODUCTIONKINETICMONTE_2007,chatterjee_OverviewSpatialMicroscopic_2007}. 
The possible cell identities represent coarse-grained regions of the system and selected chemically important groups, 
preserving microscopic morphology and active or reactive sites of interest.

Our representative physical system is a hydrated anion exchange membrane (AEM) composed of polyphenylene oxide trimethyl ammonium (PPO-TMA), a polyelectrolyte material with fixed cationic sites, and an electrolyte that contains chloride as a counter-ion for the TMA$^+$~\cite{r.varcoe_AnionexchangeMembranesElectrochemical_2014}. The structure of a PPO-TMA repeat unit is shown as the inset of Fig. \ref{fig:boxes}. The electrolyte in the hydrated AEM self-assembles into a network of connected nanoscale water channels that percolate the material and facilitate transport of anions. The self-assembled morphology of this material is also observed \textit{in silico}, where it emerges from the short-ranged interaction potentials used in MD simulation and has been validated against X-ray scattering experiments~\cite{lu_ParameterizationCoarsegrainedModel_2017}. Contemporary studies of AEMs are most interested in its chemical (or electrochemical) degradation, which occurs over the timescale of hours or days. While the effects of degradation may be studied by mutating an AEM into its degraded form, capturing the evolving breakdown of an AEM is orders of magnitude beyond the reach of MD simulation. kMC is a complementary approach that excels at modeling rare-event kinetics over relatively long time scales. Additionally, equilibrated MD snapshots can be directly converted to kMC starting configurations, an approach that has been used to project MD simulations forward through time, expanding the simulations from nanoseconds to weeks of simulated time~\cite{gadea_KineticMonteCarlo_2026}. This represents a massive extension in 
simulation time, but this MD-kMC approach is still limited in 
size 
by the MD starting configuration.

In the MD simulation, phase-separated domains and a self-assembled network of water channels emerge from short-ranged pairwise interactions, with characteristic length scales well beyond the range of the interactions themselves. Here, pairwise potentials with a cutoff on the order of $\sim 10$ \AA\ give rise to a percolated pore network whose characteristic spacing, obtained from the material's structure factor, is $\sim 3$ nm. The same local-to-global route is available in the lattice representation. If short-ranged rules can be constructed to play the role of the MD pairwise interactions, their repeated application to a randomly arranged lattice of cells should drive the emergence of a morphology with the same characteristic organization. Such an approach targets the structural signature of the equilibrated MD configuration rather than the dynamical pathway by which MD reaches it, and it requires no representation of the underlying MD dynamics.

This is an implementation of cellular automata (CA) where the lattice of cells, each in a discrete state, are updated in parallel using rules based on their local environment. From the repeated application of the proper transition rules a global behavior, in this case the AEM morphology, will emerge~\cite{wolfram_StatisticalMechanicsCellular_1983}. Further, with well-formed rules, the size of the generated lattice would not be confined to the dimensions of the corresponding MD simulation. A lattice of arbitrary size may be generated. These larger starting configurations will immediately benefit kMC simulation work, which was previously confined in length scale by MD's computational expense.

Cellular automata have proven successful in modeling microstructural evolution across diverse materials systems. Early applications focused on solidification phenomena~\cite{rappaz_ProbabilisticModellingMicrostructure_1993}, grain growth~\cite{anderson_ComputerSimulationGrain_1984}, and recrystallization~\cite{hesselbarth_SimulationRecrystallizationCellular_1991, raabe_CouplingCrystalPlasticity_2000,davies_GrowthNucleiCellular_1997} where the discrete nature of grain boundaries and crystallographic orientations naturally maps to a lattice representation. In these systems, transition rules, often inspired by Potts models or simplified Monte Carlo energetics~\cite{anderson_ComputerSimulationGrain_1984,holm_MisorientationDistributionEvolution_2001}, successfully reproduce experimentally observed microstructures. More recently, CA methods have been applied to crack propagation~\cite{silling_MeshfreeMethodBased_2005}, corrosion~\cite{dicaprio_3DCellularAutomata_2016}, and additive manufacturing~\cite{zinoviev_EvolutionGrainStructure_2016}, demonstrating the versatility of the framework when appropriate local rules can be formulated.

Despite success in crystalline materials, CA applications remain relatively uncommon in soft matter and polymer systems. The central difficulty is formulating transition rules that capture emergent behavior arising from complex, often continuous physics—chain connectivity, entanglements, and long-range correlations in polymer conformations using only local neighborhood information. Hand-crafted rules based on physical intuition may fail to reproduce the subtle interplay of interactions governing phase separation in these systems. Alternative approaches to microstructure generation include phase field methods~\cite{chen_PhaseFieldModelsMicrostructure_2002}, which excel at capturing interface dynamics but
require an appropriate representation of the underlying morphology. For complex polymer systems, constructing such a representation from the molecular structure can be challenging, particularly when molecular-scale features such as cation placement are important. More recent machine learning techniques such as generative adversarial networks~\cite{yang_MicrostructuralMaterialsDesign_2018,mosser_ReconstructionThreedimensionalPorous_2017} and variational autoencoders~\cite{chan_MachineLearningCoarse_2019}, can generate realistic microstructures but may lack direct physical interpretability and often require substantially more training data.

The core challenge of this CA approach is developing a set of proper `rules' that emulate the MD force field. Our approach uses logistic regression to drive a data-driven discovery of the effective local interactions. This method combines the computational efficiency of CA with the physical grounding of MD-derived statistics: the learned rules implicitly encode the energetics necessary for phase separation while preserving the interpretability and simplicity of local transition probabilities. With these short-ranged rules in place, applying a series of CA swap moves results in the emergence of long-range order, polymer-water phase segregation with morphological character similar to that observed in both MD simulation and the experimental system.

\section{Computational Methods}
\label{sec:methods}

\subsection{MD simulation and kMC mapping}

We obtain the equilibrated MD configuration of a hydrated AEM shown in Figure \ref{fig:boxes} by constructing a low-density simulation cell, followed by a series of compression and annealing steps. MD simulation implemented the high-resolution, coarse-grained force field FF\textsubscript{pvap}, developed for the efficient simulation of hydrated PPO-TMA membranes. For details regarding force field development, preparation and execution of the MD simulation, and model validation we refer the readers to previous work by Molinero and co-workers.~\cite{lu_ParameterizationCoarsegrainedModel_2017,barnett_PostHydrationCrosslinkingIon_2023} 

To map the MD snapshot to a kMC lattice as in Fig \ref{fig:boxes}, we simplify the system to 3 components: polymer, water, and cation, where `cation' represents the positively-charged TMA sites bound to the polymer. 
We label each lattice by $\lambda$, the number of water molecules per cation in its parent molecular configuration. Although this molecular definition of $\lambda$ is not preserved by the gridding procedure, it provides a convenient label for the hydration state of the configuration from which the lattice was constructed.

Each lattice cell has a nominal edge length of $4$~\AA{}. The realized edge is set by tiling the cubic MD box with the integer number of cells whose edge is closest to, and no smaller than, $4$~\AA\ 
For a box edge $\ell$ the lattice contains $L = \lfloor \ell/4~\text{\AA} \rfloor$ cells per side, with a realized cell edge $a = \ell/L$. The $\lambda=10$ training configuration ($\ell = 93.9$~\AA) thus gives $L=23$ and $a = 4.08$~\AA, for $\lambda=20$ ($\ell = 82.4$~\AA) it gives $L=20$ and $a = 4.12$~\AA. The identity of each cell is determined by the composition of the corresponding region of the MD configuration: `cation' if one or more TMA$^+$ is present, `polymer' if more polymer particles are present than electrolyte, else `water.'
Formally, we represent the cubic lattice as having $L$ cells per edge, $\mathcal{L} \in \{\text{polymer, water, cation}\}^{L \times L \times L}$. These selections reduce the complexity of the representation while retaining information critical to describing the state of the hydrated polyelectrolyte since the corresponding kMC model was developed to simulate degradation by cation loss. The 3-D lattice, with three possible discrete states for each cell, also fully defines the CA framework that will be used to generate new kMC starting configurations.   

\subsection{Development of CA rules}

The kMC configuration derived from an MD snapshot is used as a training data set for binary logistic regression. In this example the cubic training set has 23 cells per side and is periodic in all 3 dimensions. We divide the generation of a new lattice based on this training set into two parts: the formation of the polymer/water phases and the placement of the cations since the system is effectively a two-phase material with point defects. 

\subsubsection{Generating polymer and water domains}

\begin{figure}[h]
\includegraphics[width=3.25in]{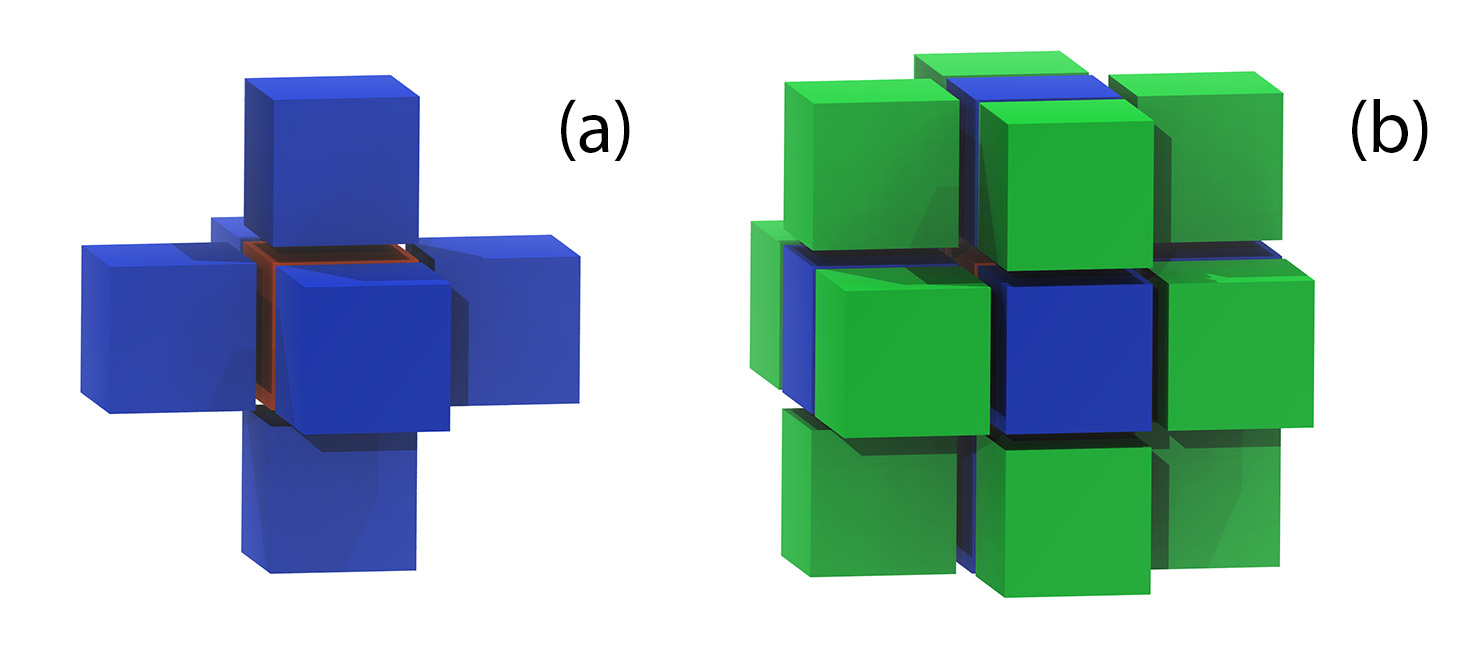}
\caption{\label{fig:neighbors} Illustrations of the first (blue cells, panel (a)) and first and second (blue, green, panel (b)) neighbor `shells' around a central lattice cell (red, partially visible in (a)) used to develop the CA rules.} 
\end{figure}

 The local neighborhood of each cell is defined to include two shells: the 6 axial neighbors at offsets $(\pm1,0,0)$, $(0,\pm1,0)$, and $(0,0,\pm1)$ and the 12 face-diagonal neighbors at $(\pm1,\pm1,0)$, $(\pm1,0,\pm1)$, and $(0,\pm1,\pm1)$, all including the possibility of wrap across periodic boundaries. For a lattice site $i$, $\mathcal{N}^{(s)}(i)$ denotes the set of neighbors in shell $s$. Illustrations of these neighbor shells are shown in Figure \ref{fig:neighbors}. Each lattice site is represented by a simple feature vector $x$, with neighbor counts $ n_y^{(s)}$ aggregated by shell resulting in $x(i) = [n_0^{(1)}(i),\; n_1^{(1)}(i),\; n_0^{(2)}(i), \; n_1^{(2)}(i)]$. The superscript $s$ indicates the neighbor shell and the subscript $y \in \{0,1\}$ denotes the class, with 0 corresponding to polymer and 1 to water. Cation sites were randomly replaced by polymer or water according to the global polymer/water ratio in the training set prior to training. Class scores for each lattice site are defined as $s_y(x) = w_y^\top x + b_y$ and the conditional probabilities are obtained by applying a softmax over the two scores,

\begin{equation}
P(y \mid x) = \frac{\exp\!\big(s_y(x)\big)}{\sum_{y' \in \{0,1\}} \exp\!\big(s_{y'}(x)\big)} \quad \text{for } y \in \{0,1\}.
\end{equation}

\noindent The parameters $\{w_y,b_y\}$ are fit by minimizing L2-regularized negative log-likelihood using the SAGA solver. 

To generate a new lattice of edge length $L$ a new lattice is initialized with the desired polymer fraction, arranged randomly. Cellular automata evolution proceeds as repeated local swap attempts between axial neighbors. For each attempted swap a random site $i$ and its random axial neighbor $j \in \mathcal{N}^{(1)}(i)$ are selected. If $y(i)=y(j)$, do nothing. Otherwise, let $a=y(i)$ and $b=y(j)$ and calculate a local plausibility score $S$ for before and after the swap,

\begin{align}
S_\text{before}=P(a \mid x(i)) + P(b \mid x(j))\\
S_\text{after}=P(b \mid x(i)) + P(a \mid x(j)).
\end{align}

\noindent If $S_\text{after} > S_\text{before}$ then accept the swap. 

Each CA cycle consists of $N/5$ swap attempts, where $N$ is the number of lattice cells, with the initiating site and its axial partner drawn uniformly at random. Evolving for 500 cycles therefore corresponds to 100 proposed swaps per lattice cell. This cutoff of 500 cycles is set by the approach to a stationary morphology: the accepted-swap rate falls to a low background within a few hundred cycles, after which swaps continue to flicker at domain interfaces without appreciable further coarsening. Extending the evolution from 500 to 1500 cycles changes the characteristic domain size by less than a cell edge (see Supporting Information). Because the acceptance criterion is greedy, each run relaxes into one of many arrangements consistent with the learned rules, selected by the random initial configuration; individual lattices therefore differ in detail while sharing the same characteristic structure.

\subsubsection{Cation placement}

Cations are placed after the polymer and water domains have been generated. The local neighborhood is defined as above using the 6 axial neighbors and the 12 face-diagonal neighbors. For each site $i$ the 6-dimensional feature vector aggregates neighbor counts of all three classes by shell,

\begin{equation}
x(i) = \big[n_0^{(1)}(i),\; n_1^{(1)}(i),\; n_2^{(1)}(i),\; n_0^{(2)}(i),\; n_1^{(2)}(i),\; n_2^{(2)}(i)\big],
\end{equation}

\noindent where $y \in \{0,1,2\}$ denotes polymer, water, and cation respectively. A multinomial logistic regression is trained on the labeled training lattice, with class scores $s_y(x) = w_y^\top x + b_y$ and softmax posterior
\[
P(y \mid x) \;=\; \frac{\exp\!\big(s_y(x)\big)}{\sum_{y'\in\{0,1,2\}} \exp\!\big(s_{y'}(x)\big)} \quad \text{for } y\in\{0,1,2\}.
\]


From the training lattice, we also compute a normalized histogram of first-shell cation hydration, $h_k = \Pr\big(n_1^{(1)}(i)=k \mid y(i)=2\big)$ for $k=0,\dots,6$. Given a target number of cations $N_\text{cat}$ in the generated lattice, the desired counts per hydration bin are $d_k = N_\text{cat}\, h_k$.

Cations are then placed sequentially into the polymer–water lattice until $N_\text{cat}$ is reached. To reduce cost, rolling first-shell maps $n_0^{(1)}$, $n_1^{(1)}$, and $n_2^{(1)}$ are maintained and updated incrementally after each placement. At each iteration:

1) Sample a candidate subset $C$ from non-cation sites $i$ with $y(i)\in\{0,1\}$. For each $i\in C$, compute $p_\text{cat}(i) = P(y=2 \mid x(i))$ from the trained classifier and record $k(i) = n_1^{(1)}(i)$, the number of first-shell water neighbors.

2) Let $c_k$ be the current number of placed cations observed with $k$ first-shell water neighbors. Define a normalized histogram deficit reward
\[
r_\text{hist}(i) \;=\; \frac{\max\!\big(d_{k(i)} - c_{k(i)},\, 0\big)}{N_\text{cat}}.
\]

3) Form a convex combination score that trades off matching the global hydration profile and the local classifier posterior,
\[
S(i) \;=\; w\, r_\text{hist}(i) \;+\; (1-w)\, p_\text{cat}(i),
\]
with histogram weight $w\in[0,1]$ (default $w=0.5$).

4) Select one site from $C$ to place a cation via Boltzmann sampling,
\[
\pi(i) \;=\; \frac{\exp\!\big(\alpha\, S(i)\big)}{\sum_{j\in C} \exp\!\big(\alpha\, S(j)\big)},
\]
where $\alpha>0$ controls exploitation versus exploration (default $\alpha=10$). 

5) Place the cation at the chosen site $i$, set $y(i)\leftarrow 2$, increment $c_{k(i)}$, and update the first-shell neighbor maps on $\mathcal{N}^{(1)}(j)$: for each neighbor $j\in \mathcal{N}^{(1)}(i)$, increase $n_2^{(1)}(j)$ by 1 and decrease $n_0^{(1)}(j)$ or $n_1^{(1)}(j)$ by 1 according to the original phase at $i$. Second-shell counts in $x(i)$ are recomputed on demand for candidate evaluation.

\begin{figure}[h]
\includegraphics[width=6.25in]{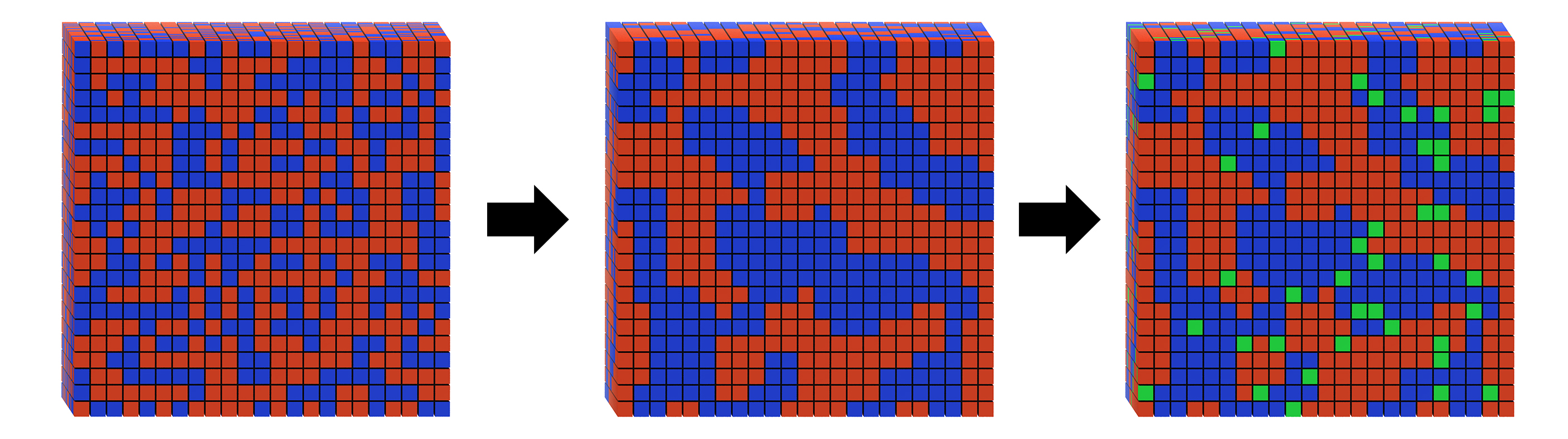}
\caption{\label{fig:ca} A random configuration of water(blue) and polymer (red) cells (left panel) is arranged into water/polymer domains by 500 cycles of CA swaps (center panel.)  Cation sites are placed in a second step, with positions based on both local neighborhood and a global histogram of cation hydration (right panel.)} 
\end{figure}

Fig. \ref{fig:ca} graphically summarizes the CA-based generation of AEM lattices.

The process begins with a randomly arranged lattice of polymer and water cells, whose ratio equals that of the MD-derived training set. Repeated local swaps then coarsen this random arrangement into phase-separated domains, reaching a stationary morphology within the 500 cycles specified above. A second stage places cationic sites using the local rules and cation-neighbor statistics described above, continuing until the total number of cationic sites matches that of the parent configuration.

This 
procedure ensures that the generated lattice matches both the global composition and the local environment statistics of the training set, as encoded by the ML-derived CA rules. 
Our framework is flexible and can generate new configurations at water content levels which differ from the training set (demonstrated later in this work) and this method may be extended to other microstructures by learning the corresponding local rules from a high-fidelity training structure.

\subsection{Structure-factor window and comparison metrics}
\label{sec:metrics}
 
We quantify the agreement between generated and reference lattices using two complementary structural metrics. The ordinal cation-neighbor distributions probe the short-range environment, directly reflecting the local coordination statistics learned in the CA rules, while the spherically averaged structure factor $S_\alpha(q)$ probes long-range order that develops beyond the spatial extent of the training neighbors. We report the neighbor distributions as normalized histograms and $S_\alpha(q)$ as a curve over its reliable wavevector range. For quantitative comparison, we additionally summarize each metric using scalar descriptors.

We compute $S_\alpha(q)$ for the polymer and water phases using the species-resolved occupation variable $\rho_\alpha(\mathbf{r}) = \delta_{y(\mathbf{r}),\alpha} - \langle\delta_{y(\mathbf{r}),\alpha}\rangle$, where $y(\mathbf{r})\in\{0,1,2\}$ is the cell identity at lattice site $\mathbf{r}$ and the subtracted mean removes the $\mathbf{q}=0$ contribution. The structure factor is

\begin{equation}
    S_\alpha(\mathbf{q}) = \frac{1}{N}\left|\sum_{\mathbf{r}} \rho_\alpha(\mathbf{r})\, e^{i\mathbf{q}\cdot\mathbf{r}}\right|^2,
\end{equation}

\noindent where $N$ is the total number of lattice cells and $\mathbf{q}$ runs over the discrete  Fourier modes of the periodic lattice, $q_\mu = 2\pi n_\mu / (L_\mu a)$, with $a$ the cell edge length, $L_\mu$ the number of cells along direction $\mu$, and  $n_\mu \in \{-L_\mu/2,\ldots,L_\mu/2-1\}$. A spherically averaged $S_\alpha(q)$ is  obtained by binning all $\mathbf{q}$ by magnitude $q = |\mathbf{q}|$.

Because the lattice is finite and periodic, the spherically averaged structure factor $S_\alpha(q)$ is a reliable probe of bulk morphology only over an intermediate band of wavevectors. Below the minimum-image wavevector $q_\mathrm{MI} = 2\pi/(\ell/2) = 4\pi/\ell$, corresponding to features larger than half the box edge, $S_\alpha(q)$ is dominated by the discrete, sparsely populated low-order reciprocal-lattice shells and cannot separate genuine long-wavelength structure from correlations imposed by the periodic boundary. Above the voxel Nyquist wavevector $q_\mathrm{Ny} = \pi/a$, $S_\alpha(q)$ probes sub-cell length scales that the lattice discretization does not resolve. We therefore restrict quantitative comparison of $S_\alpha(q)$ to the reliable window $q_\mathrm{MI} \le q \le q_\mathrm{Ny}$; for the $\lambda=10$ lattice ($\ell = 93.9$~\AA, $a = 4.08$~\AA) this window is $0.13 \le q \le 0.77$~\AA$^{-1}$.

For the ordinal cation-neighbor distributions we report the signed shift in mean neighbor count, $\Delta\langle k\rangle = \langle k\rangle_\mathrm{gen} - \langle k\rangle_\mathrm{ref}$, and the Wasserstein-1 (earth-mover) distance $W_1$ between the two normalized histograms. For $S_\alpha(q)$ we report the peak position $q^\ast$ (refined by parabolic interpolation about the maximum bin), a normalized $L_1$ curve discrepancy over the reliable window, $D_1 = \big(\sum_q |S_\mathrm{gen}(q) - S_\mathrm{ref}(q)|\big) / \big(\sum_q S_\mathrm{ref}(q)\big)$, and the fraction of spectral weight falling below $q_\mathrm{MI}$.

For both structural metrics, we report a measure of statistical significance (whether a difference is resolvable above run-to-run scatter) and a measure of effect size (whether it is large compared with that scatter). For the kMC trajectories we compare ensemble-mean curves by their root-mean-square difference normalized by the single-run, run-to-run standard deviation, $\mathrm{RMSD}/\sigma$; a ratio below unity indicates a mean difference smaller than the intrinsic run-to-run spread. Where replicates share random-number seeds across the two lattices we additionally report a paired effect size $|d_z|$ and the $p$-value of a sign-flip permutation test over the seed-matched differences; where they do not, an unpaired permutation test is used. Uncertainties on ensemble quantities are reported as $\pm$ one standard deviation across independently generated lattices.

\section{Results and Discussion}

\subsection{CA-generated lattice structure and morphology}

Our initial study used a $\lambda=10$ configuration taken from a previous computational study of PPO-TMA~\cite{erimban_DegradationAnionExchange_2024,erimban_EtherCleavageDecreases_2025}. This MD simulation cell consisted of 34875 total coarse-grained particles and had an edge length of 9.4 nm and was converted into a $23 \times 23 \times 23$ lattice, each lattice cell corresponding to 4.08 \AA. This lattice was used as a training set for the approach outlined in Section \ref{sec:methods}. The training and generation of a new lattice based on CA rules was performed on Lawrence Livermore National Laboratory's Dane supercomputer. Each Dane node is based on Intel Sapphire Rapids processor with 56 cores per socket, 2 sockets per node, and 256 GB DDR5 memory. 
While running on a single CPU core, the training and generation of an independent new lattice of side 23 took approximately 12 seconds of wall time, which equates to $\sim 3 \times 10^{-5}$ node-hours.

\begin{figure}[h]
\includegraphics[width=3.25in]{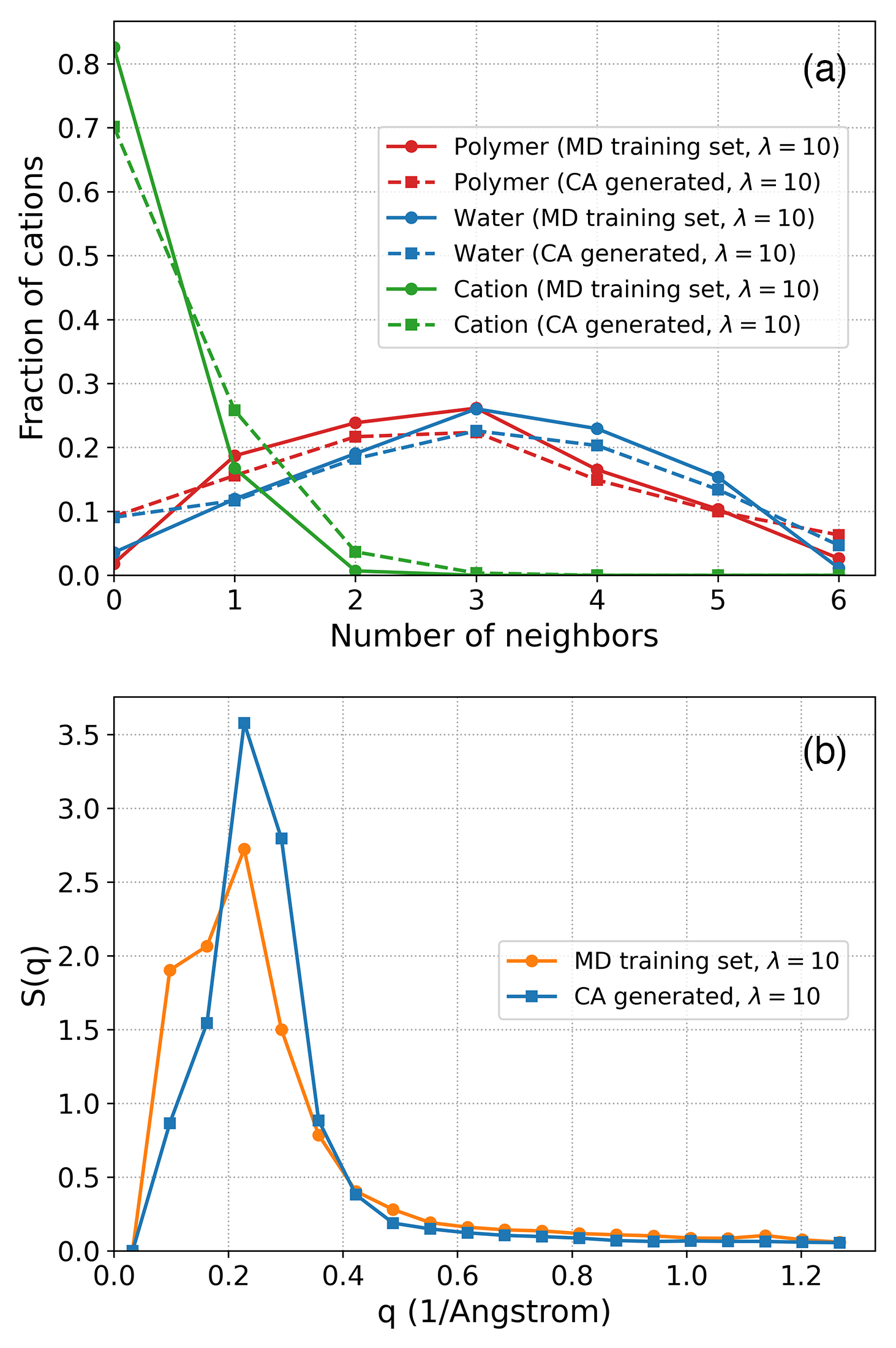}
\caption{\label{fig:10hist} (a) Probability distributions of cation local neighborhoods for MD training sets (solid curves) and distributions from the respective CA-generated $\lambda=10$ AEMs (dashed curves.) (b) Corresponding water and polymer structure factors for the $\lambda=10$ case.} 
\end{figure}

The cation-neighbor distribution describes the local environments around cationic sites and is directly connected to the CA training strategy. Fig \ref{fig:10hist}(a) compares this distribution for the training set and generated lattices. Across ten independently generated $\lambda = 10$ lattices, the CA approach reproduces the training set's neighbor statistics closely, with mean-neighbor shifts relative to MD of $\Delta\langle k\rangle = +0.01 \pm 0.03$ (polymer), $-0.12 \pm 0.04$ (water), and $+0.11 \pm 0.03$ (cation) neighbors (Wasserstein-1 distances $0.20$, $0.22$, and $0.11$ neighbors, respectively). The water-neighbor distribution, which the placement step targets explicitly through the cation hydration histogram, is reproduced to within ${\sim}0.1$~neighbor of the MD mean. The cation-cation coordination is not targeted by the placement objective and emerges slightly elevated relative to MD; consistent with its unconstrained origin, this shift is small, systematic, and highly reproducible across runs.

The CA reproduces the characteristic long-range spacing of the water-channel network, with both the MD-derived and generated lattices showing a correlation peak near $0.2$~\AA$^{-1}$ (Fig.~\ref{fig:10hist}(b)). This peak corresponds to a ${\sim}30$~\AA{} characteristic spacing and lies well within the reliable $0.13 \le q \le 0.77$~\AA$^{-1}$ window (Sec.~\ref{sec:metrics}).
Within this window the MD and CA peaks agree in position to within the wavevector bin width ($\Delta q = 0.065$~\AA$^{-1}$) for both water and polymer, and the curves differ by a normalized discrepancy $D_1 = 0.33 \pm 0.04$ (water) and $0.28 \pm 0.04$ (polymer), dominated by a higher CA peak amplitude.

The MD reference additionally exhibits a shoulder near $0.08$~\AA$^{-1}$. This corresponds to a ${\sim}79$~\AA{} length that exceeds the $47$~\AA{} half-box length and is resolved by only the first one-to-two reciprocal-lattice shells. This feature lies below $q_\mathrm{MI}$, in the regime where $S_\alpha(q)$ is not a reliable measure of bulk structure: whether it reflects genuine long-wavelength heterogeneity or correlations induced by the periodic boundary, the two cannot be separated in a cell of this size. The CA-generated lattice does not reproduce this shoulder, and because the feature lies outside the trustworthy window its absence is not a morphological deficiency. Quantitatively, the MD places $8.6\%$ (water) and $7.3\%$ (polymer) of its spectral weight below $q_\mathrm{MI}$, of which the CA reproduces roughly a third ($2.8 \pm 0.7\%$ and $3.0 \pm 0.8\%$); the CA concentrates the remaining ${\sim}5$ percentage points into the physical peak, accounting for its higher amplitude. This is consistent with local transition rules that reproduce the dominant correlation length but cannot generate the box-scale heterogeneity underlying the shoulder.

\subsection{kMC results}

\begin{figure}[h]
\includegraphics[width=3.25in]{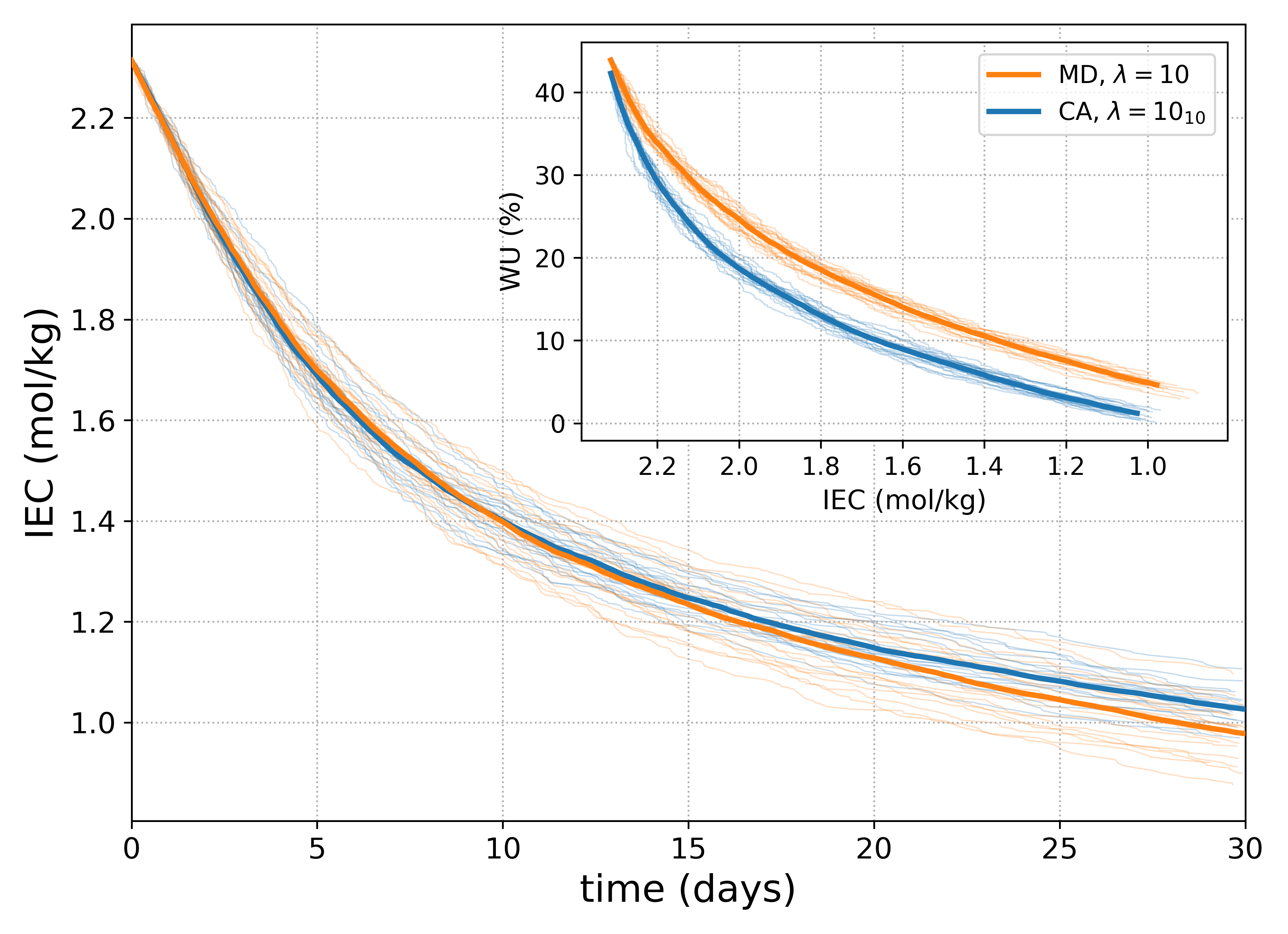}
\caption{\label{fig:10kMC} IEC versus time for membrane degradation simulated by kMC. Inset: Corresponding Water uptake versus IEC values from the same kMC simulations. Bold curves represent the averages of 20 kMC replicate runs, each individual run shown a thin line of corresponding color.} 
\end{figure}

These CA-generated starting configurations are designed to reproduce local ordering and long-range morphology in order to serve as inputs for larger length scale kinetic Monte Carlo simulations of membrane degradation. To this end, we now compare the simulated degradation of a lattice directly derived from MD simulation with that of the lattice generated by our cellular automata approach, shown in Figure \ref{fig:10kMC}. Both starting configurations consist of $23 \times 23 \times 23$ cells and represent the same geometric size simulation cell. We perform kMC simulation using the same parameters in Reference ~\cite{gadea_KineticMonteCarlo_2026}. In brief, this approach simulates E2 degradation under 100\% relative humidity, with kinetic parameters taken from experimental studies ~\cite{willdorf-cohen_AlkalineStabilityAnionExchange_2023,muller_PracticalExSituTechnique_2020}. We emphasize that the present work does not seek to quantitatively reproduce experimentally-observed degradation of a polyelectrolyte membrane. Our aim is to show that these CA-generated membranes perform similarly as MD-derived configurations in kMC simulation while requiring significantly less computational resources to prepare than an MD-derived lattice.

Figure~\ref{fig:10kMC} shows the degradation of the membranes over 30 days of simulated time. Here we define degradation as the loss of cationic sites due to an E2 elimination reaction whose kinetics are a strong function of local cation hydration as detailed in Reference ~\cite{gadea_KineticMonteCarlo_2026}. Ion exchange capacity (IEC), the number of moles of cationic sites per kg of dry polyelectrolyte, is a convenient metric to quantify this chemical degradation. In each panel the bold curves are averages over 20 kMC replicates and the thin curves are the individual runs, included to convey the run-to-run variance of the kMC approach at this lattice size; the 20 replicates use identical random-number seeds across the MD-derived and CA-generated lattices, so the two ensembles can be compared run-by-run.
 
The two approaches produce statistically indistinguishable degradation kinetics: over a 30-day window the CA-generated and MD-derived average IEC$(t)$ curves differ by an RMS of $0.024$~mol\,kg$^{-1}$, only $0.50\times$ the run-to-run standard deviation ($0.048$~mol\,kg$^{-1}$), and a paired permutation test over the 20 replicates with identical randomization seeds resolves no systematic lattice effect ($p = 0.13$). That the two agree so closely is intuitive, since cation placement is trained on the local hydration of the cation and the degradation rate is most sensitive to the number of neighboring water cells.
 
The resulting water uptake, however, reveals a small but clear systematic
difference. The inset of Fig.~\ref{fig:10kMC} shows water uptake (WU), the mass percent of water in the hydrated membrane, as a function of decreasing IEC. 

Both lattices begin with very similar cell populations but the CA-generated membrane retains less water as degradation proceeds, with a mean offset of $\Delta\mathrm{WU} \approx 4.9\%$ (an absolute difference on the ${\sim}42\%$ baseline of the hydrated membrane), which is large relative to the run-to-run spread ($\mathrm{RMSD}/\bar\sigma = 5.6$, paired $|d_z| \approx 3.7$, $p < 10^{-3}$).
 
The kMC implementation spawns an auxiliary particle (AP) upon cation loss, and the evolution of these APs is a strong function of the polymer/water morphology. The two metrics therefore report on complementary length scales: IEC versus time on the short-range chemistry of cation loss, and WU versus IEC on the longer-range morphology through which the APs evolve. 
That the degradation kinetics are indistinguishable while the water uptake is not places the residual difference between the two lattices at the longer length scale, consistent with the small low-$q$ residual in the structure factor noted above. The first-shell hydration of the two lattices agrees closely enough $\Delta\langle k_w\rangle = -0.12$) that the kinetics, despite their strong hydration dependence, are unaffected. 

\subsection{Extrapolation to other hydration levels}

\begin{figure}[h]
\includegraphics[width=3.25in]{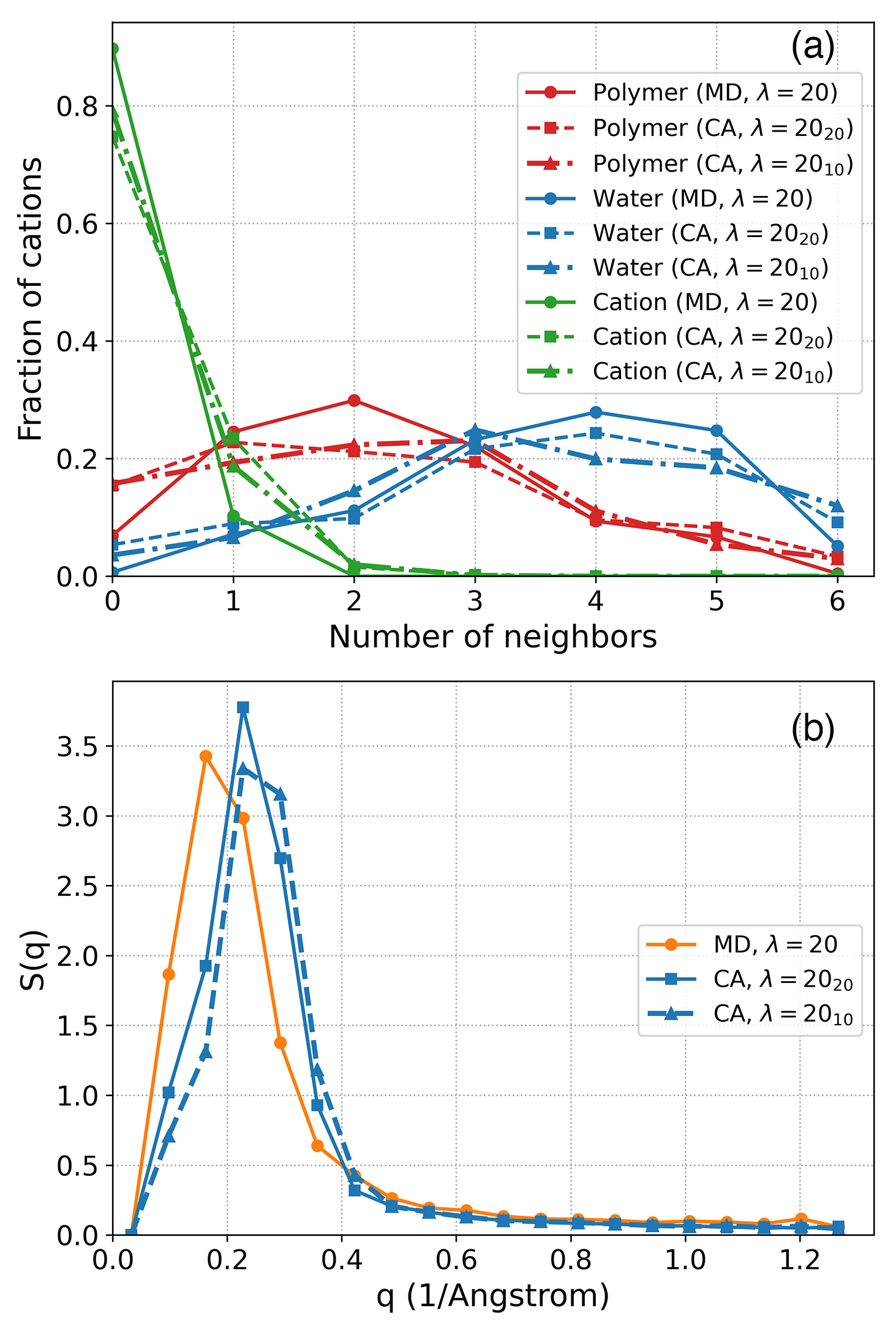}
\caption{\label{fig:20hist} (a) Probability distributions for cation neighborhoods in the $\lambda=20$ MD ``ground-truth'' lattice (solid lines) and corresponding CA-generated $\lambda=20$ AEMs using different training data. (b) Corresponding structure factors for the $\lambda=20$ training set and generated AEMs.} 
\end{figure}

The training set and generated lattice considered above both represent a membrane with hydration $\lambda=10$. In ion exchange membranes $\lambda$ 
is a function of IEC, relative humidity, the identity of the cationic site, and other factors. In simulation studies of these materials researchers are interested 
in simulating properties at several hydration levels, with each $\lambda$ requiring the preparation of a new MD starting configuration. If CA rules can be transferred between hydration levels, this presents another potential opportunity to reduce 
computational expense: extrapolation along $\lambda$. In this approach, we consider generating CA-based lattices with a range of lambda, with CA rules derived from  the same $\lambda=10$, MD-derived configuration. Fig. \ref{fig:20hist} shows an overview of these extrapolation results. We use the nomenclature $\lambda = X_Y$, meaning `$\lambda = X$ trained on $\lambda = Y$,' where $X$ is the target $\lambda$ value of the CA-generated AEM lattice and $Y$ is the $\lambda$ value of the MD simulation cell used to create the CA rules. For example, in the $\lambda = 20_{10}$ case, a $\lambda = 10$ MD configuration was used 
to create CA rules 
which were then used to generate a $\lambda = 20$ CA lattice. 
In Figure \ref{fig:20hist}, we compare the MD-based ``ground truth'' for $\lambda=20$ with the ``properly trained'' $\lambda=20_{20}$ and the extrapolated $\lambda=20_{10}$ case, in which the CA rules were trained using a $\lambda=10$ MD configuration.

To a first approximation, the probability histograms of cation neighborhood populations in Fig.~\ref{fig:20hist}(a) look similar for the MD, $20_{20}$, and $20_{10}$ lattices. The number of water molecules solvating the cationic site (blue curves) is the most important factor for the subsequent kMC simulations
since the degradation rate in this model depends sensitively on cation hydration. Both CA-generated lattices reproduce the MD cation hydration 
well: the properly trained $20_{20}$ lattice agrees with the $\lambda=20$ MD ground truth to $\Delta\langle k_w\rangle = -0.16$ neighbors, 
while the extrapolated $20_{10}$ lattice (generated from rules learned from the $\lambda=10$ training lattice) is within $\Delta\langle k_w\rangle = -0.11$. That extrapolation reproduces the target hydration about as well as direct training is consistent with the central premise of this method: the learned transition rules are local, so the first-shell environment they generate transfers across hydration levels much as local structure does in the underlying MD simulations. The number of polymer neighbors in $20_{10}$ peaks at a slightly higher value, also intuitive since the ratio of water cells to polymer cells is a fixed input to the CA generation step. We note that the present comparison is a single extrapolation of modest range ($\lambda=10\rightarrow20$) within the percolated AEM domain. A systematic study of the limits of extrapolation, e.g. toward membrane dehydration and percolation loss, is left to future work since our immediate focus is system scale-up at fixed hydration.


Panel~\ref{fig:20hist}(b) shows the 
long-range ordering that emerges in the generated $\lambda=20$ lattices. The $\lambda=20$ MD ground truth peaks lowest, at $q^\ast \approx 0.18$~\AA$^{-1}$; both CA-generated lattices peak at slightly higher $q$ (marginally smaller channel spacing), $q^\ast \approx 0.24$~\AA$^{-1}$ for the properly trained $20_{20}$ lattice and $\approx 0.26$~\AA$^{-1}$ for the extrapolated $20_{10}$. All three peaks lie within the reliable window, and the differences are at the scale of the wavevector bin width ($\Delta q = 0.065$~\AA$^{-1}$). The properly trained lattice matches the ground-truth peak to $|\Delta q^\ast| \approx 0.06$~\AA$^{-1}$ and the extrapolated lattice to $\approx 0.07$~\AA$^{-1}$. Both CA lattices thus reproduce the dominant correlation length of the $\lambda=20$ membrane to within about one bin, with the properly trained lattice marginally closer, consistent with the near-equivalent cation hydration of panel (a).

\begin{figure}[h]
\includegraphics[width=3.25in]{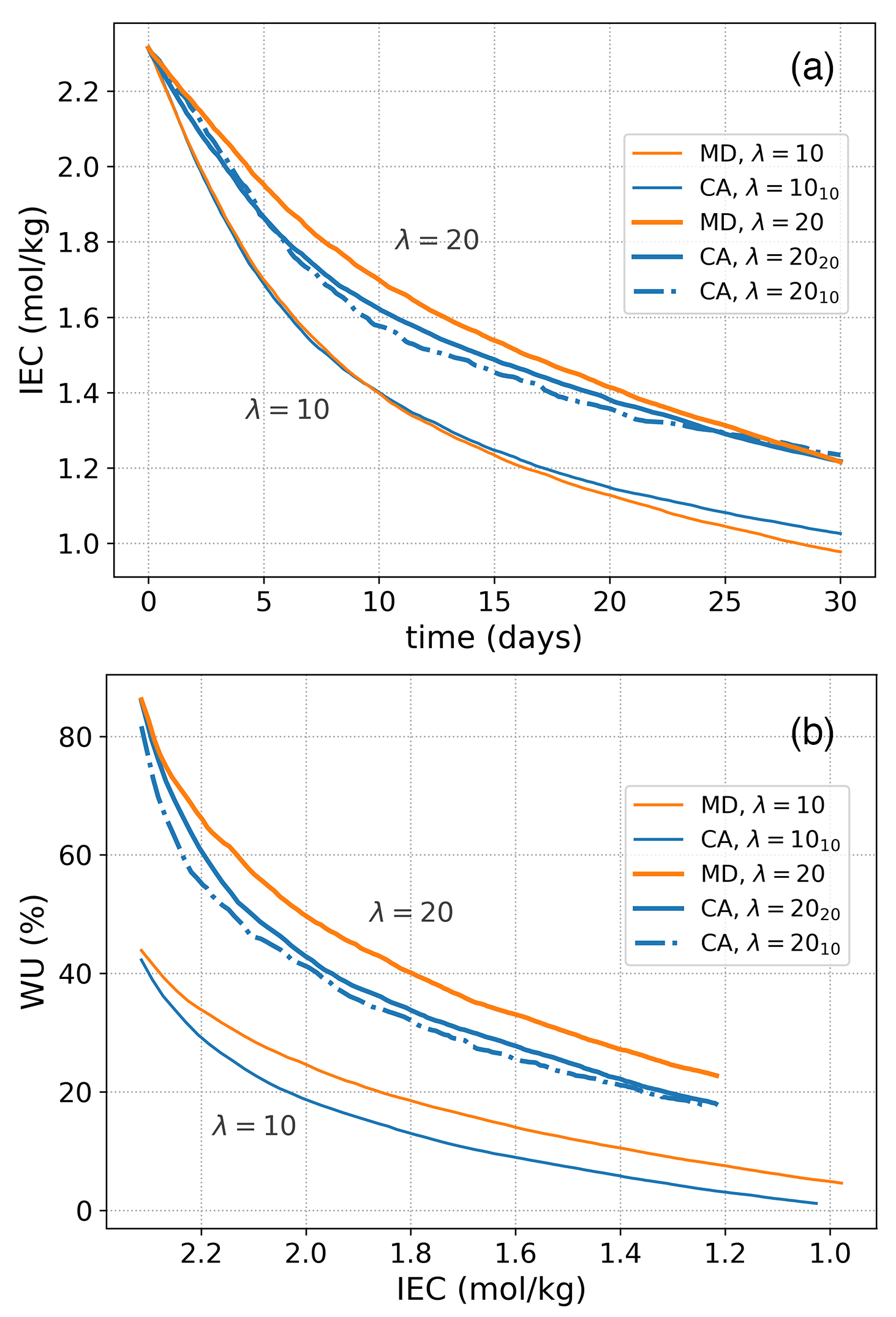}
\caption{\label{fig:20kMC} (a) Simulated degradation results for CA-generated and MD-mapped lattices. (b) WU vs IEC values from the same set of kMC simulations. Both $\lambda=10$ and $\lambda=20$ conditions are shown.} 
\end{figure}

Just as Fig.~\ref{fig:20hist} assessed the effect of hydration extrapolation on the generated lattice structure, Fig.~\ref{fig:20kMC} evaluates its impact on subsequent kMC simulations. The figure compares the $\lambda=20$ MD-derived ``ground truth'' lattice with the properly trained $\lambda=20_{20}$ lattice and the extrapolated $\lambda=20_{10}$ lattice. Panel (a) shows the evolution of IEC during degradation. As expected, the properly trained $\lambda=20_{20}$ lattice reproduces the MD degradation kinetics more closely than the extrapolated $\lambda=20_{10}$ lattice. This behavior is consistent with the cation hydration distributions in Fig.~\ref{fig:20hist}(a): Because the $\lambda=20_{10}$ lattice inherits a first-shell environment from the $\lambda=10$ training configuration, it contains fewer water neighbors around each cation than the MD and $\lambda=20_{20}$ lattices, leading to faster degradation in the kMC model.

Panel ~\ref{fig:20kMC}(b) shows the evolution of water uptake during degradation. Unlike the IEC curves, which primarily reflect the local environment governing cation loss, the water uptake depends on the longer-range morphology through which the auxiliary particles introduced by the kMC model evolve. The $\lambda=20_{20}$ lattice again more closely reproduces the MD behavior, whereas the larger deviation of $\lambda=20_{10}$ is consistent with its greater long-range structural discrepancy relative to the MD lattice. Together with the low-$q$ structure factor results discussed above, this suggests that hydration extrapolation preserves local environments more effectively than long-range morphology.


\clearpage

\subsection{Scale up}

\begin{figure}[h]
\includegraphics[width=3.25in]{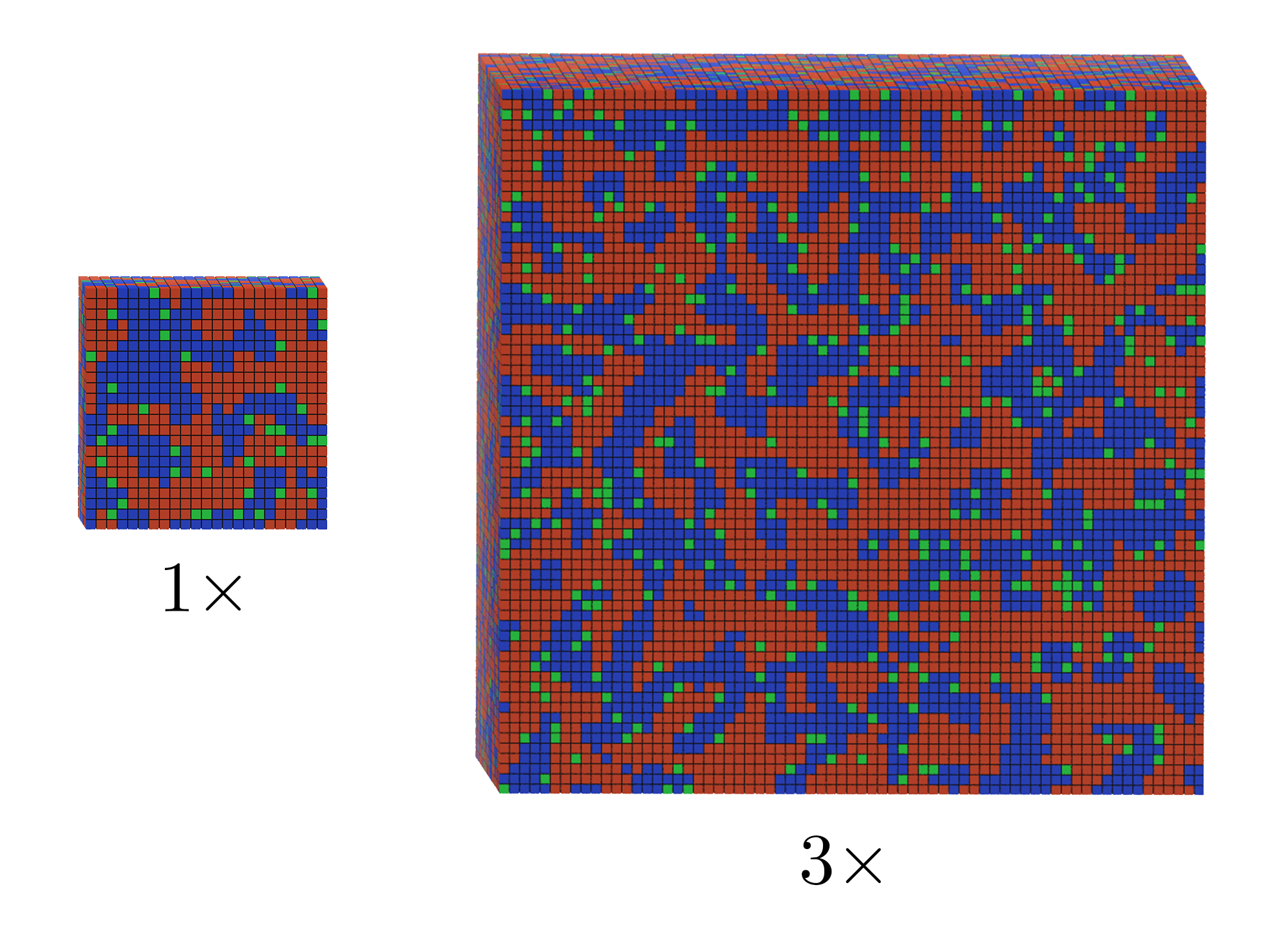}
\caption{\label{fig:3box} Representative CA-generated $1\times$ and $3\times$-scaled up AEM configurations.} 
\end{figure}

Having calibrated this CA approach against a molecular dynamics `ground truth,' we now consider scale-up. Enabling simulation at system sizes that fine-grained simulations cannot directly reach, the regime where multi-scale modeling actually lives, is the core motivation of this work. In related studies of coarse-grained PPO-TMA, MD system size was increased to study cell size effects and related phenomena ~\cite{lu_EffectPolymerArchitecture_2019}. 
A common approach to study size effects is to double the simulation-cell edge length, which increases the simulation volume, and therefore the number of particles, by nearly an order of magnitude, substantially increasing computational cost. Inspired by this standard $2 \times$ approach, we consider scale up of CA-generated lattices by increasing the lattice cube edge $3 \times$, corresponding to a volumetric (or particle number) increase of $27 \times$, to ensure that we enter a regime which would have been difficult for our MD simulations to reach. This scale-up involves training on the $23 \times 23 \times 23$ $\lambda=10$ lattice described earlier, followed by generation of the larger lattice. Figure \ref{fig:3box} illustrates the scale-up by showing representative configurations of the original and $3\times$ lattices side by side, highlighting the substantial increase in system size relative to the MD training system.

Increasing the system volume by $27\times$ increased the compute time by a factor of approximately $49 \times$, from $\sim 12$ s for the $1 \times$ ($23^3$) lattice to $\sim 600$ s (about 10 minutes) for the $3 \times$ lattice on a single CPU core. That the time grows somewhat faster than the volume reflects an empirical scaling of wall time as roughly $N^{1.16}$ in the number of lattice cells $N$, with the cost split between the CA swap evolution and the cation-placement step.
We emphasize that the code developed for this initial work is serial and not fully optimized, neither for speed nor memory use, but report resource use to provide a baseline both for scaling system size and for comparison to MD simulation. 

\begin{figure}[h]
\includegraphics[width=3.25in]{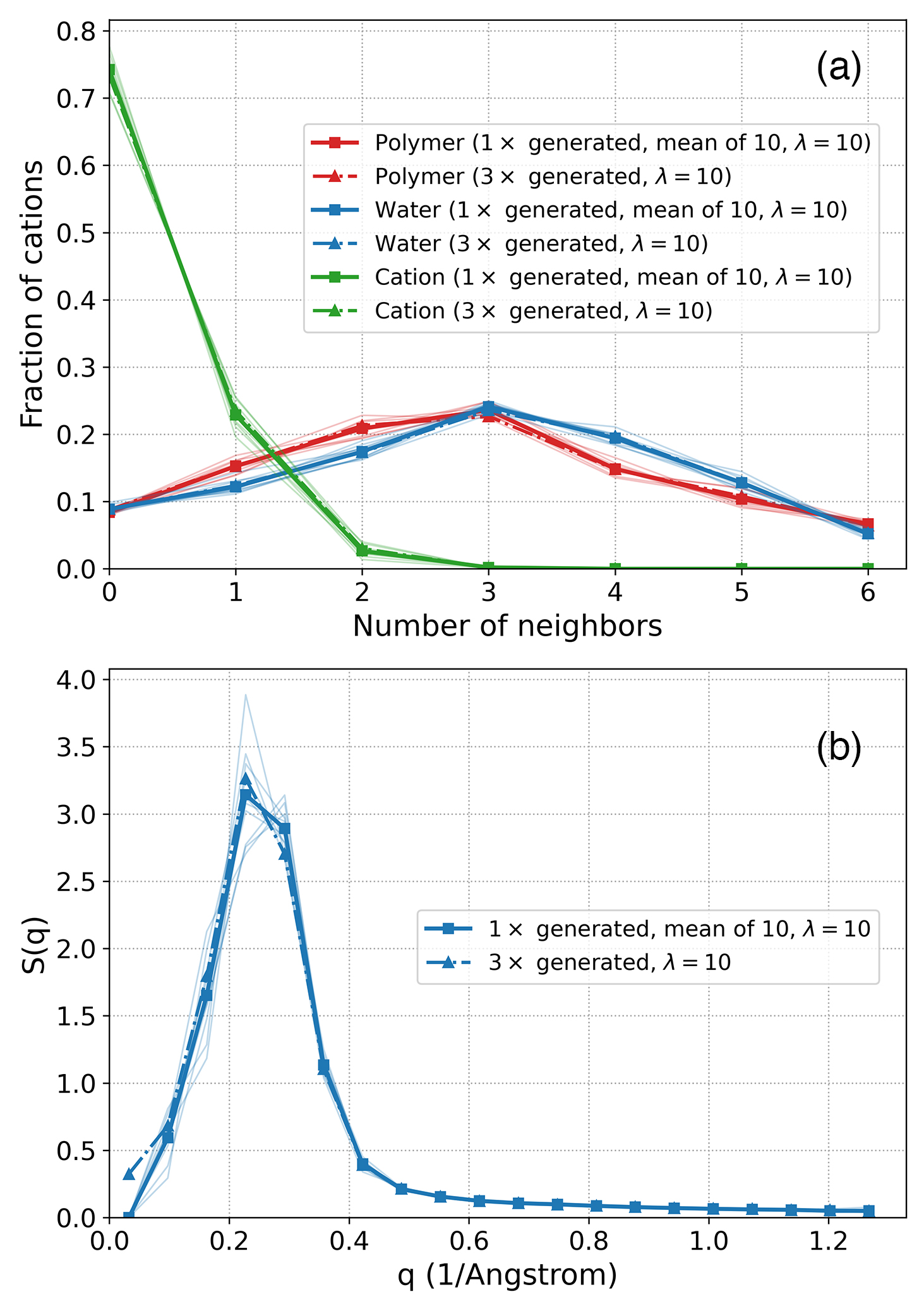}
\caption{\label{fig:3xhist} (a) Cation neighborhood probability histograms for CA-generated $\lambda=10$ lattices. Solid curves represent lattices of the same size as the MD-based training set and dashed curves are for generated lattices with $3\times$ larger edge lengths. (b) Corresponding structure factors $S(q)$ for the $1 \times$ and $3 \times$ lattices. Faint curves represent the 10 individual $1 \times$ lattices from which the mean is obtained.} 
\end{figure}

The main concern in this scale up is whether the CA method of generating starting configurations is sensitive to the size of the generated lattice: Do the learned CA rules restrict generated lattices to the size of the training data lattice? Figure \ref{fig:3xhist} directly compares lattices generated at the same size as the MD-derived training set, `$1 \times$' or $23 \times 23 \times 23$ cells, with a $3 \times$ larger edge length ($27 \times$ greater volume) generated cell. Solid curves indicate the average of 10 independently generated $1 \times$ lattices, the faint lines are the individual $1 \times$ lattices and the dashed curves represent a $3 \times$ scaled-up lattice. Figure \ref{fig:3xhist} (a) shows cation neighbor histograms and (b) shows structure factors $S(q)$. The $3 \times$ lattice reproduces the $1 \times$ neighbor statistics and structure factor, confirming that the CA generation is size-consistent. The mean cation hydration shift relative to the MD reference is unchanged between sizes ($\Delta\langle k_w\rangle = -0.115 \pm 0.039$ at $1 \times$ versus $-0.110 \pm 0.009$ at $3 \times$), as are the polymer and cation-neighbor shifts and the structure factor peak position ($q^\ast = 0.25$ versus $0.24$~\AA$^{-1}$, within the wavevector bin width). The small systematic biases identified at $1 \times$ (cation under-hydration and slight cation-cation excess) persist unchanged at $3 \times$, establishing them as size-stable properties of the method rather than finite-size artifacts.
 
The larger lattice is markedly more reproducible: the run-to-run standard deviation contracts by roughly a factor of four to eight on scaling up (e.g. the water $\Delta\langle k_w\rangle$ standard deviation falls from $0.039$ to $0.009$, and the $S(q)$ peak-position standard deviation from $0.011$ to $0.001$~\AA$^{-1}$), consistent with self-averaging over the $27 \times$ larger volume. Finally, the fraction of spectral weight below the minimum image limit increases modestly at $3 \times$ ($0.043$ versus $0.028$ for water): the larger box resolves genuine low-$q$ modes that the $1 \times$ cell could not sample, rather than amplifying a simulation box-size artifact, and this weight remains small in absolute terms.

\begin{figure}[h]
\includegraphics[width=3.25in]{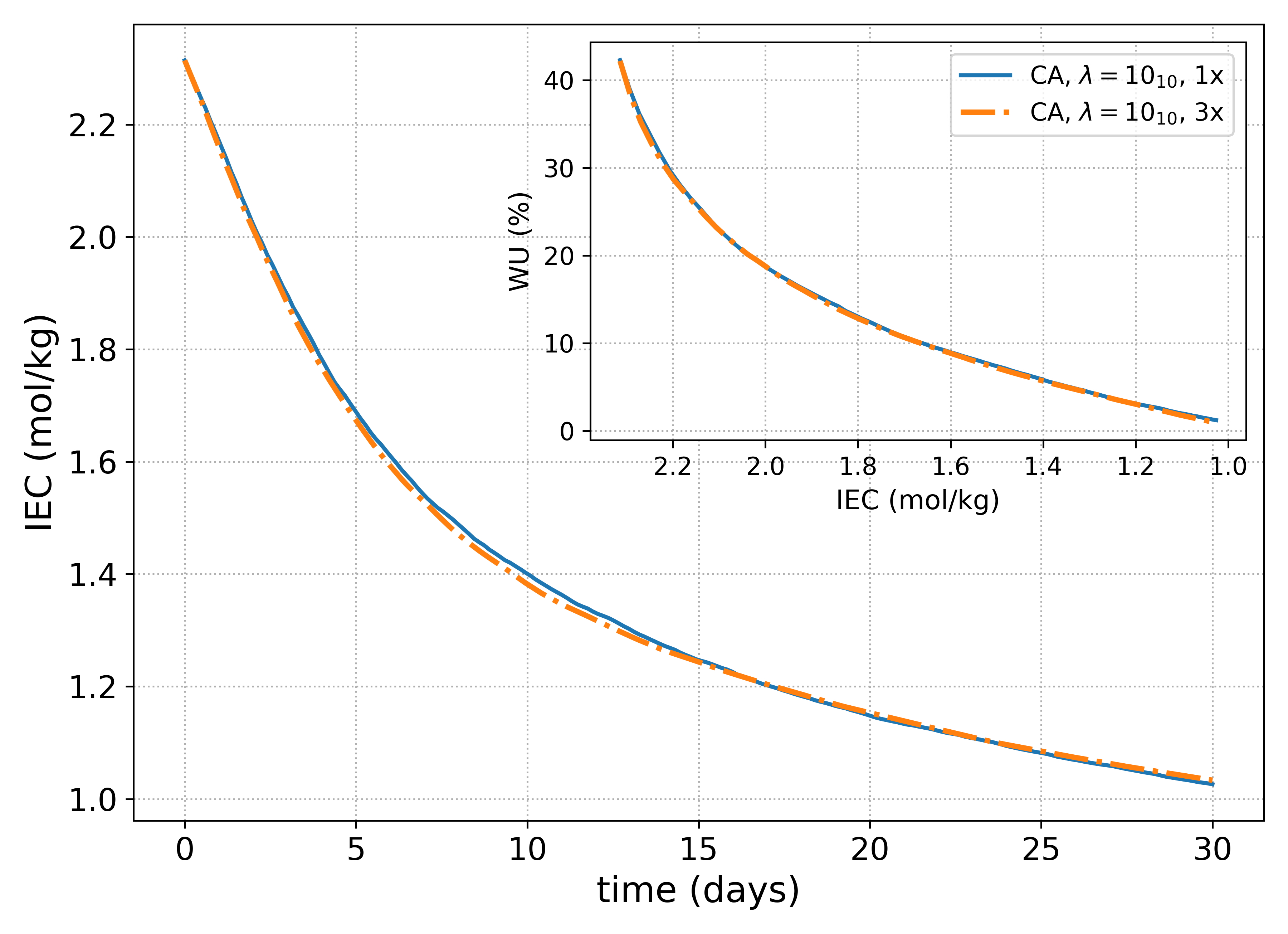}
\caption{\label{fig:3x-kMC} kMC simulated degradation of CA-generated lattices whose size match the training set and $3\times$ larger edge length CA lattices. Inset: Corresponding WU versus IEC values.} 
\end{figure}

Lastly, and most importantly, we must confirm that larger simulation cells retain the ability to function as kMC starting configurations. If simulated degradation of a larger cell does not agree with the same simulation of the corresponding smaller lattice, there exists some feature or behavior which prevents extension of this approach to the generation of larger starting configurations. In this case we are not concerned with comparison to a `ground truth' MD simulation. Figure \ref{fig:3x-kMC} overlays kMC simulated degradation results for the two lattices. IEC vs. time curves for the $\lambda=10$, $23 \times 23 \times 23$ (blue, solid) and $\lambda=10$, $3 \times$ larger cell (orange, dashed) overlay very closely, with the difference in average for a given time being less than the observed run-to-run variance for a given lattice size. The inset of Figure \ref{fig:3x-kMC} shows similar agreement for WU vs. IEC. These results confirm that, with appropriately defined CA rules, the generation of arbitrarily large starting configurations is a valid application of this lattice generation method. 

\section{Conclusions}

We have presented a cellular automata approach for generating lattice starting configurations suitable for mesoscale simulation, with the goal of decoupling the accessible length scale of a lattice-based method from the size of the fine-grained simulation used to parameterize it. Using a single equilibrated MD snapshot of a hydrated PPO-TMA anion exchange membrane as a training set, we learned local transition rules via logistic regression over a two-shell neighborhood and applied them as a sequence of local swap moves on a randomly initialized lattice. Cation placement was handled in a second step that combines the learned local classifier with a global hydration histogram, ensuring that both short-range environment and macroscopic distribution match the training data.

The resulting CA-generated lattices reproduce the cation neighborhood statistics and the spherically averaged structure factor of the MD-derived training set, capturing both the short-range chemistry and the emergent polymer–water phase separation that underlies anion transport in these materials. More importantly, when used as starting configurations for kMC simulations of E2 degradation, the CA-generated lattices yield IEC versus time and water uptake versus IEC curves that are quantitatively consistent with those obtained from MD-derived starting configurations. The method is computationally inexpensive: 
training and generation of a benchmark-sized ($23^3$) lattice required about 12 seconds on a single CPU core of LLNL's Dane system, and scaling to a $27 \times$ larger volume increased the wall time to roughly 10 minutes. kMC degradation results for the $3 \times$ edge-length lattice agree with the benchmark within the run-to-run variance, confirming that the learned rules generalize across system size and that the approach can supply starting configurations at length scales beyond the practical reach of the originating MD simulation.


Several natural extensions are worth noting. Rules trained at one hydration level extrapolated to another with only modest loss of fidelity: generating a $\lambda=20$ lattice from $\lambda=10$-trained rules reproduced both the long-range morphology and the cation hydration of the $\lambda=20$ ground truth about as well as rules trained directly at $\lambda=20$. This transferability follows from the local character of the learned rules and reinforces the view of the CA method as a coarse-grained mimic of the local physics that MD encodes. Mapping the limits of this extrapolation, such as toward membrane dehydration and loss of water-channel percolation, is a natural direction for future work. 

More broadly, the framework as presented is restricted to a two- or three-component, single-site-per-cell representation; extending it to multicomponent systems, vector-valued cell states, or anisotropic neighborhoods would broaden its applicability to other soft matter and polymer problems where local rules are  difficult to hand-craft but training data from higher-fidelity simulation is available.
We view the present work as proof of principle that data-driven CA rules can serve as a lightweight, interpretable bridge between particle-based and lattice-based simulation, and we anticipate that the same philosophy will prove useful well beyond the AEM systems considered here.

\section{Acknowledgements}

This work was supported by the U.S. Department of Energy (DOE) Industrial Technologies Office (ITO) under Funding Opportunity Announcement DE-FOA-0002804, with additional support from a Cooperative Research and Development Agreement (CRADA TC02455) among Lawrence Livermore National Laboratory (LLNL), TotalEnergies, Siemens Energy, and Ionomr Innovations, Inc. Work at LLNL was performed under the auspices of the DOE under contract DE-AC52-07NA27344 and is approved for unlimited release as LLNL-JRNL-2020696. 

\clearpage


%


\clearpage

\setcounter{figure}{0}
\setcounter{table}{0}
\setcounter{equation}{0}
\setcounter{section}{0}
\renewcommand{\thefigure}{S\arabic{figure}}
\renewcommand{\thetable}{S\arabic{table}}
\renewcommand{\theequation}{S\arabic{equation}}
\renewcommand{\thesection}{S\arabic{section}}

\begin{center}
{\large\bfseries Supporting Information for\\
``Bridging simulation length scales with cellular automata''}
\end{center}
\vspace{1em}

\section{Convergence of cellular automata method}

\begin{figure}[h]
\includegraphics[width=3.25in]{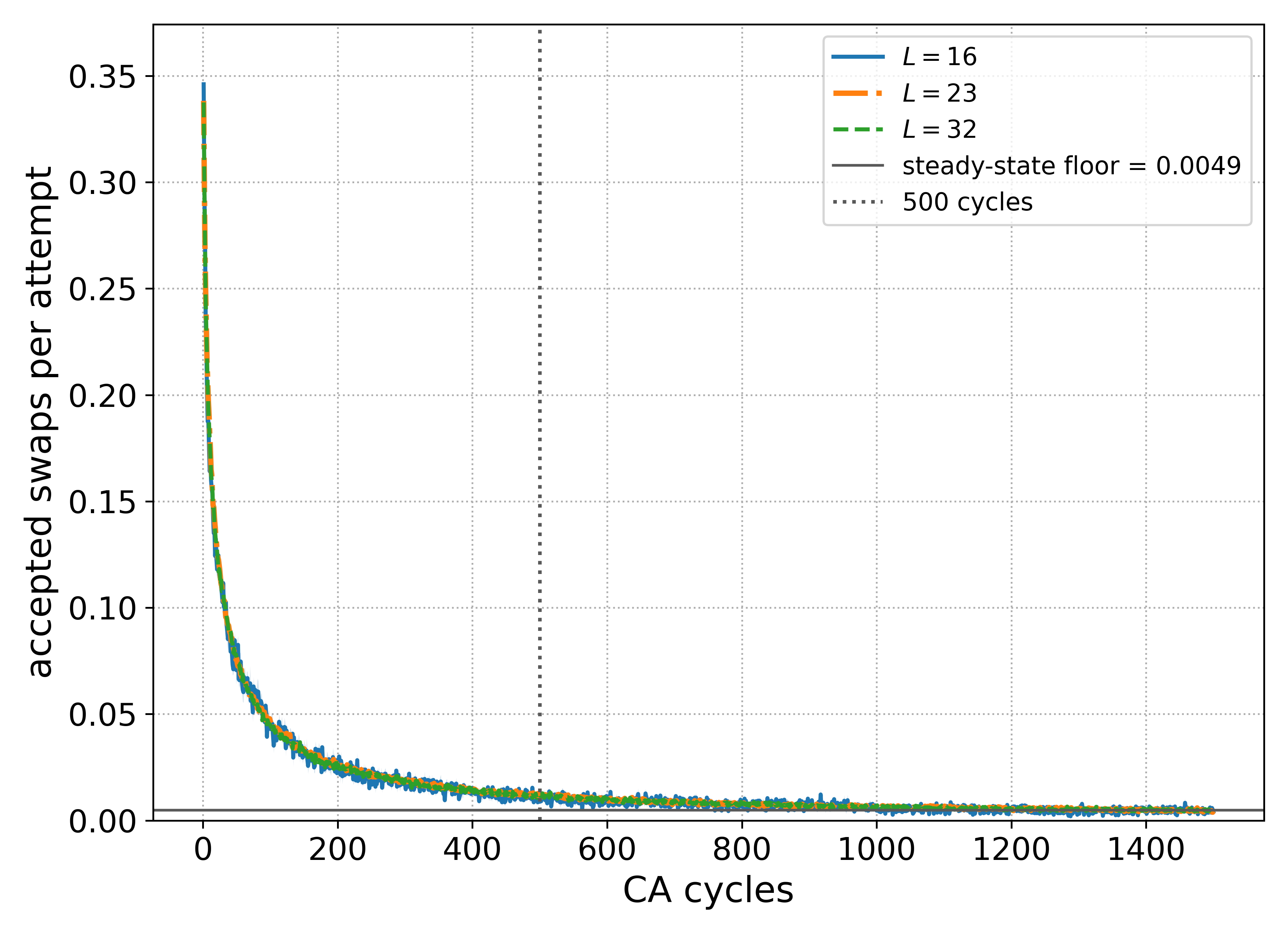}
\caption{\label{fig:acceptance} The fraction of accepted CA swaps per attempt is shown for three different lattice sizes. Our selected cutoff of 500 cycles is indicated by the vertical dashed line and the non-zero steady-state selection floor is shown as a horizontal solid line.} 
\end{figure}

Our approach applies 500 cycles of cellular automata swap moves on a cubic lattice whose cells are randomly initialized to one of two possible states. In each cycle, swap rules are applied to $N/5$ of the sites so that an average of 100 swap attempts is applied to each cell. During this stage a two-phase morphology emerges. Swap-rule acceptance rates settle by $\sim260$ CA steps and we selected 500 cycles, about twice this number of cycles, as the point to select a morphology for the second phase, cation placement. Figure \ref{fig:acceptance} shows acceptance rate per CA swap attempt for runs extended to $3\times$ our nominal endpoint.

\begin{figure}[h]
\includegraphics[width=3.25in]{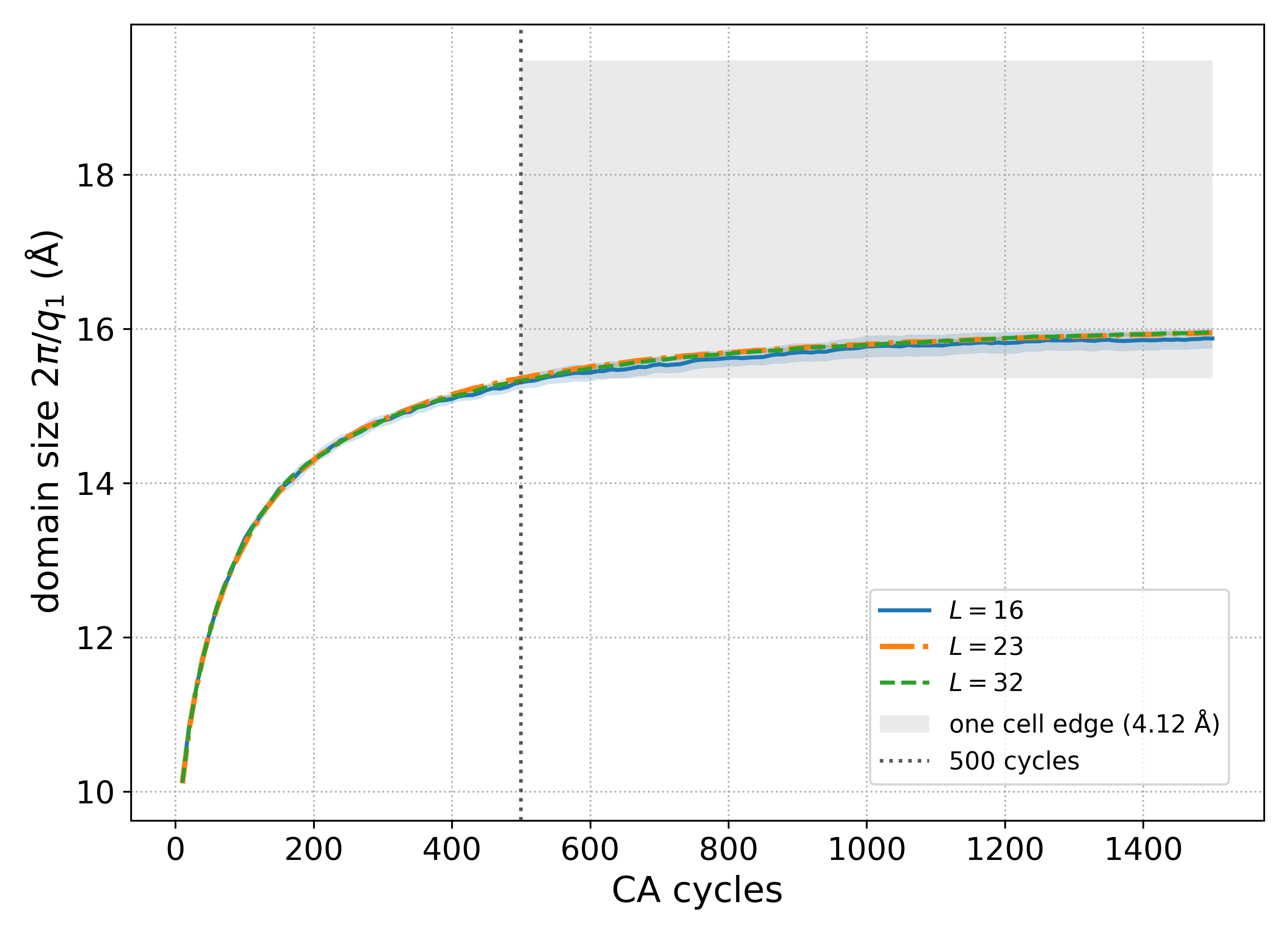}
\caption{\label{fig:domain} Domain size versus cycle for three different cell sizes. The shaded region indicated the size of a single lattice cell; changes in the plateau region beyond 500 cycles are a fraction of a single lattice cell.} 
\end{figure}

Extending runs by a factor of 3, to 1500 cycles, shows that structural changes between 500 and 1500 steps is well below the length of one lattice cell edge ($\sim 0.14$). The evolution of domain size as a function of cycle is shown in Figure \ref{fig:domain}. We anticipate that this method is extensible to other physical systems of interest and expect that the number of CA cycles is not universal and is a function of the CA swap rules native to each system of study.

\clearpage

\end{document}